\documentstyle[12pt,epsfig]{article}

\def\beq{\begin{equation}}
\def\eeq{\end{equation}}
\def\bea{\begin{eqnarray}}
\def\eea{\end{eqnarray}}
\def\bq{\begin{quote}}
\def\eq{\end{quote}}
\def\ra{\rightarrow}

\def\la{\lambda}

\def\eps{\epsilon}
\def\bq{\begin{quote}}
\def\eq{\end{quote}}
\def\ra{\rightarrow}

\begin{document}
\vspace*{-1.5cm}
\thispagestyle{empty}
\begin{flushright}
hep-ph/yymmxxx\\
IOA-TH/98-03\\
NTUA-70/98
\end{flushright}

\vspace{2.5cm}
\begin{center}
{\large
{\bf Lepton Flavour Violation  in }\\
{\bf Unified Models with $U(1)$-Family Symmetries$^*$}\\

\vspace{.8cm}
G.K. Leontaris$^a$ and N.D. Tracas$^b$
}

\vspace{0.8cm}
{\sl
$^a$Theoretical Physics Division, Ioannina University, GR-451 10 Ioannina,
Greece}\\
{\sl
$^b$Physics Department, National Technical University, 157 80 Athens,
Greece}

\vspace{1cm}
{\bf Abstract}
\end{center}

\vspace{.5cm}\noindent
Lepton flavour non-conserving processes are examined in the context
of unified models with $U(1)$-family symmetries which reproduce
successfully the low-energy hierarchy of the fermion mass spectrum and
the Kobayashi - Maskawa  mixing. These models usually imply mixing
effects in the supersymmetric scalar sector. We construct the fermion
and scalar mass matrices  in two viable models, and calculate the
mixing effects on the $\mu\ra e\gamma$, $\mu\ra 3 e$ and $\tau\ra 
\mu\gamma$ rare decays.  The relevant constraints on the
sparticle mass spectrum as well as the role of various MSSM parameters
are  discussed.

\vfill
\hrule\vspace{.3cm}  \noindent{\small
$^*$Research supported in part by TMR contract ERBFMRX-CT96-0090,
$\Pi E NE\Delta$-1170/95 and $\Pi E  NE\Delta$-15815/95\\

\vspace{.3cm}\noindent
February 1998
}
\newpage

One of the most dramatic consequences of supersymmetric extensions of the 
Standard Model (SM) is the appearance of new sources of flavour violations
\cite{oldflav,a,j}.
Supersymmetric partners of gauge, fermion and scalar fields generate new
types of flavour violating diagrams at the one-loop level, which enhance
considerably the various rare processes.

 Flavour non-conserving processes may still be relatively suppressed
if the matrices of the supersymmetric partners of fermions, i.e. those
of scalar quarks and scalar leptons, are diagonal in flavour space. It
is widely believed however, that a realistic spectrum for the fermion
mass matrices can be obtained when additional symmetries discriminate
between the various families of the known fermion fields. Such
symmetries imply also a non-trivial structure for the corresponding
scalar mass matrices. Rare processes, being sensitive to these
changes, usually lead to hard violation of flavour.

In this work, we compute the branching ratios for lepton flavour
violating decays in realistic models whose fermion and scalar mass
textures are obtained by $U(1)$-family ($U(1)_f$-) symmetries. In our
analysis we choose both, symmetric and non-symmetric fermion textures
by appropriate selection of the $U(1)_f$ fermion charges. The choice
of the lepton sector to test the predictability, and possibly the
viability, of $U(1)_f$-models is ideal as there are no large
uncertainties (unlike the case of the quark sector where large
ambiguities enter due to poor knowledge of hadronic matrix elements)
in the calculations. In a previous work \cite{LT}, we have given some
estimates for the $\mu\ra e\gamma$ decay is the context of a simple
$U(1)_f$-model in the case of small $\tan\beta$ regime. Here, we
extend our previous analysis and examine cases for large and small
$\tan\beta$ and various values of the gravitino and gaugino masses.
We use two-loop analysis for gauge couplings and take into account
threshold effects to calculate the sparticle spectrum used to
construct the relevant scalar and fermion mass matrices entering the
above processes.  We find that the non-observation of lepton flavour
violating processes put rather strong lower limits on the sparticle
mass spectrum, in particular when $\tan\beta$ is large.

A wide class of models, which naturally incorporate flavour
non-diagonal scalar mass matrices, arises in string scenarios where the
usual gauge symmetry is accompanied by a number of $U(1)$ factors, the
latter playing the role of family symmetries. The fermion mass
textures of the above models are dictated by the particular charges of
the particles under the $U(1)_f$ symmetries, the specific flat
direction which has been chosen as well as the string selection rules
and other string symmetries. In general, there are only few tree-level
couplings in the superpotential (usually only those  responsible for
the top, bottom and tau masses), while all other fermion mass
entries are supposed to be generated by higher non-renormalizable
terms.

Once the flat directions and the $U(1)_f$ charges of a particular
model have been fixed, the scalar mass matrix structure may also be
easily computed through the K\"ahler function
\begin{equation}
{\cal G}={\cal K}+\log|{\cal W}|^2
\end{equation}
where ${\cal W}$ is the superpotential and ${\cal K}$ has the general form
\begin{equation}
{\cal K}=-\log(S+\bar S)-\sum h_n\log(T_n+\bar T_n)+
              Z_{i\bar j}(T_n,\bar T_n)Q_i\bar Q_j+\cdots
\end{equation}
with $Q_i$ being the matter fields, $S$ the dilaton,
whereas $T_n$ are the other moduli fields.
The scalar mass matrices are determined by $Z_{i\bar j}$ and ${\cal W}$. 
The form of the $Z_{i\bar j}$ function is dictated by the modular symmetries
and depends on the moduli and the modular weights of the fields. Thus,
at the tree level, the diagonal terms are the only non-zero entries in
the scalar mass matrices. Higher order terms allowed by the symmetries
of the specific model fill in the non-diagonal entries. In what
follows, we will explore the flavour violating processes in two
different models which give realistic fermion mass spectrum. In
particular, we will calculate the $\mu\ra e\gamma$, $\tau\ra \mu
\gamma$ and $\mu\rightarrow 3 e$ processes in the context of 
supersymmetric models whose low energy theory is the MSSM
model augmented by a $U(1)$ family symmetry. One of them
is using a charge assignment where symmetric mass matrices appear,
while the other assumes $U(1)$-charges which lead to non-symmetric
textures.

We start with some preliminary remarks about the sources of flavour
violations and set our notation and conventions. After the breaking of
some possible unified symmetry to that of the Standard Model, the Yukawa
interaction, which appear in the superpotential and violate lepton
flavour, is
\beq
{\cal W}= {e^c}^T\la_e\ell H +\cdots
\eeq
where $\ell$ is the left lepton doublet, $e$ is the right singlet
lepton, $H$ is the higgs doublet and $\lambda_e$ represents the Yukawa
coupling matrix in flavour space. In addition, soft supersymmetry
breaking terms generate  mass matrices for the charged slepton fields,
denoted by $\tilde m_\ell$, $\tilde m_{e_R}$. The Yukawa and soft
scalar mass-squared matrices are diagonalized by unitary
transformations
\bea
\la_e &=& V_R^*\la_e^{\delta}V_L^{\dagger}
\\
\tilde{m}^2_\ell = U_\ell \left({\tilde m^2_\ell}\right)^{\delta}
 U_\ell^{\dagger},
\quad &\hbox{and }&
\tilde{m}^2_{e_R} = U_{e_R} \left({\tilde m^2_{e_R}}\right)^{\delta} 
                                                      U_{e_R}^{\dagger}
\eea
where $\delta$ denotes diagonal. The lepton mass eigenstates ($eig$)
are related to the weak eigenstates ($w$) by
\bea
\ell_{eig}=V_L^{\dagger}\ell_w ,&{e^c}_{eig}=V_R^{\dagger}{e^c}_w
\eea

The charged slepton mass-squared matrix is a $6\times 6$ matrix, 
built up from the the two $3\times 3$  left $\tilde{m}^2_\ell$ and right
$\tilde{m}^2_{e_R}$ soft ones, as well as the off-diagonal
submatrix which  has the form
\begin{equation}
{\left(m_{LR}^{\ell}\right)}^2=m_{\ell}(\mu\tan\beta +A_{\ell})
\label{offd}
\end{equation}
where $m_\ell=\la_e v \cos\beta/\sqrt{2}$ is the charged lepton mass
matrix, $A_\ell$ and $\mu$ are the trilinear and higgs mixing
parameters in the superpotential, $\tan\beta$ the ratio of the two
higgs vev's and $v=246$GeV.

To calculate the mixing effects in the amplitudes, we work
in the basis where the fermion mass matrix is diagonal,
\bea
{\left({{e^c}^T}\right)}_wm_{\ell}{\ell}_w&=& 
{\left({{e^c}^T}\right)}_wV_R^*m^{\delta}_{\ell}
V_L^{\dagger}{\ell}_w=
{(e^c)}^T_{eig}m_\ell^\delta \ell_{eig}
\eea
In this basis, the off-diagonal term (\ref{offd}) is written
\begin{equation}
m_{\ell}(\mu\tan\beta +A_{\ell}) =
V_R^*m_{\ell}^\delta(\mu\tan\beta +A_{\ell})V_L^\dagger
\end{equation}
This defines uniquely the scalars in the basis where
the fermions are in their mass-eigenstates
\bea
\tilde{e}_R^{*'}&=&
\left(V_R^{\dagger}\left({\tilde{e}_R}\right)\right)^*_w=
V_R^T\left(\tilde e_R\right)^*_w\\
\tilde{\ell}'&=&V_L^{\dagger}{\tilde\ell}_w.
\eea
The soft terms for right and left charged sleptons must be written
in the same basis. The 6$\times$6 matrix then takes the form
\beq
\left(
\begin{array}{cc}
V_L^{\dagger}\tilde{m}^2_{\ell}V_L & 
                       ((A_\ell+\mu\tan\beta)m_{\ell}^{\delta})^\dagger\\
(A_\ell+\mu\tan\beta)m_{\ell}^{\delta}  
                        & V_R^T\tilde{m}^2_{e_R}V_R^*
\end{array}
\right).
\label{6b6}
\eeq

A well known result in the context of the non-supersymmetric standard
model is the conservation of lepton flavour in the case of zero
neutrino masses, while in the case of massive, non-degenerate
 neutrinos, the amount of 
lepton flavour violation is proportional to the factor $\Delta
m^2_\nu/M^2_W$~\cite{P..}, which highly suppresses all relevant
 processes. When
supersymmetry enters the game, the whole scene changes  completely.
Even in the absence of right handed neutrinos, flavour violations
could occur via the exchange of supersymmetric particles. A large
number of new parameters (sparticle masses, mixing angles, e.t.c.)
appear in the calculations, enlarging therefore the parameter space and
making difficult the consistency of the predicted branching ratios with 
the experimental bounds. However, in the context of unification and low
energy phenomenology scenarios, these processes can provide useful
constraints on the parameter space.

We will briefly present the minimum number of inputs necessary to
determine all low energy parameters entering in a lepton flavour
violating process. In the context of supersymmetric unified models, we
assume a universality condition for the scalar masses at the
unification scale $M_U$. The general formula, at this scale, is $\tilde
m^2_i(M_{U})=(1+q_i)m^2_{3/2}$, where $m_{3/2}$ is the gravitino
mass, $i$ is a flavour index and $q_i$ is the modular weight of the
corresponding field. This tree-level contribution is flavour diagonal.
Non-diagonal terms are expected to appear through non-renormalizable
terms with an expense of an extra parameter $\epsilon$, namely
$\epsilon=\langle\phi\rangle/M$ where $\langle\phi\rangle$ is a
singlet field vev and $M$ a Planck-scale mass. The magnitude of the
vev $\langle\phi\rangle$
 can be fit from the fermion sector. A crucial role is also played
by the gaugino soft masses, the trilinear soft parameter $A$ as well
as the Yukawa couplings $\lambda_t$ and $\lambda_b$ at the 
unification scale $M_U$
(we assume that $\lambda_b(M_U)=\lambda_\tau(M_U)$).
In the minimal scenario, the gaugino masses at the unification scale
are determined in terms of the universal mass parameter $m_{1/2}$. 
Thus, at $M_U$, we use a minimum set of parameters,
namely ($m_{1/2}, m_{3/2}, \mu, A, \la_{t},\la_{b}$), together with
the value of the common coupling $\alpha_U$ and the unification scale
$M_U$ in such a way that after the renormalization group running we
obtain a consistent set of all low energy measured quantities. For any
acceptable such set, we calculate the branching ratios of the flavour
violating processes.

{}Figure 1 shows the one-loop diagrams relevant to the $\mu\ra
e\gamma$ process. The corresponding $\tau\ra \mu\gamma$-decay is
represented by an analogous set of graphs. $\mu\ra 3 e$ proceeds
through the decay of the (now virtual) photon to an electron-positron
pair. There are also box-diagrams contributing to this process, they
are however relatively suppressed.

\begin{figure}
\begin{center}
\epsfig{file=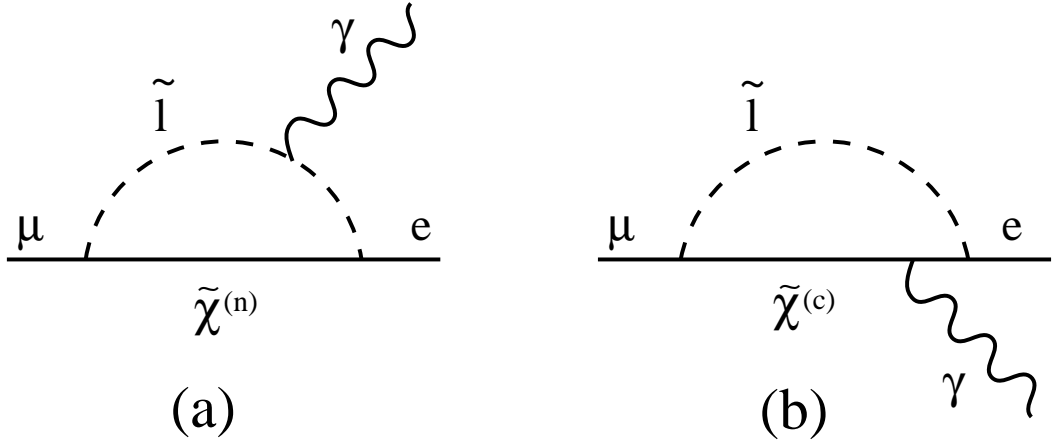,width=14cm}
\end{center}
\caption{The generic Feynman diagrams for the $\mu\ra e\gamma$
decay. $\tilde l$ stands for charged slepton (a) or sneutrino (b), while
$\tilde\chi ^{(n)}$ and $\tilde\chi ^{(c)}$ represent  
neutralinos and charginos respectively.}
\label{figure1}
\end{figure}

The electromagnetic current operator between two lepton  states $l_i$
and $l_j$ is given in general by
\begin{eqnarray}
{\cal T}_\la &=& \langle l_i|(p-q)|{\cal J}_\la|l_j(p)\rangle\nonumber\\
{  }&=&{\bar u_i}(p-q)
      \{ m_j i\sigma_{\la\beta}q^\beta 
               \left(A^L_MP_L+A^R_MP_R\right)+\nonumber\\
{  }&&\quad\quad    (q^2\gamma_\la-q_\la\gamma\cdot q)
               \left(A^L_EP_L+A^R_EP_R\right)
      \} u_j(p)
\label{general}
\end{eqnarray}
where $q$ is the photon momentum. The $A_M$'s and $A_E$'s have
contributions from neutralino-charged slepton ($n$) and
chargino-sneutrino ($c$) exchange
\begin{equation}
A_M^{L,R}=A_{M(n)}^{L,R}+A_{M(c)}^{L,R},\quad
A_E^{L,R}=A_{E(n)}^{L,R}+A_{E(c)}^{L,R}
\label{ampl}
\end{equation}
The amplitude of the process is then proportional to ${\cal T}_\la
\epsilon^\la$ where $\epsilon^\la$ is the photon polarization vector.
The easiest way to determine the loop momentum integral contribution
to the $A$'s is to search, in the corresponding diagram, {}for terms
proportional to $(p\cdot\epsilon)$ and $(q\cdot\epsilon)$. The
coefficient of the former is proportional to the momentum integral
contribution to the $\sigma_{\la\beta}$ term in (\ref{general}), while
the coefficient of the latter is proportional to the difference of the
momentum integral contribution  between the $\sigma_{\la\beta}$ and
the $(q^2\gamma_\la-q_\la\gamma\cdot q)$ terms. Defining the ratio
$x=M^2/m^2$, where $M$ is the chargino (neutralino)  mass and $m$ the
sneutrino (charged slepton) mass, the following functions appear in
the $A_M$ term
\begin{equation}
\begin{array}{ll}
A_{M(n)}:\quad&\frac{1}{6(1-x)^4}(1-6x+3x^2+2x^3-6x^2\log x) \quad\hbox{and}\\
        &\frac{1}{(1-x)^3}(1-x^2+2x\log x)\frac{M}{m_{l_j}}\\
A_{M(c)}:\quad&\frac{1}{6(1-x)^4}(2+3x-6x^2+x^3+6x\log x)\quad \hbox{and}\\
        &\frac{1}{(1-x)^3}(-3+4x-x^2-2\log x)\frac{M}{m_{l_j}}
\end{array}
\end{equation}
where $m_{l_j}$ is the mass of the $l_j$ lepton, while for the $A_E$ we
have
\begin{equation}
\begin{array}{ll}
A_{E(n)}:\quad&\frac{1}{(1-x)^4}(2-9x+18x^2-18x^3+6x^3\log x) \\
A_{E(c)}:\quad&\frac{1}{(1-x)^4}(16-45x+36x^2-7x^3+6(2-3x)\log x)
\end{array}
\end{equation}
Notice the lack of terms proportional to the gaugino mass $M$ which cancel.
The Branching Ratio (BR) of the decay $l_j\ra l_i+\gamma$ is given by
\[
BR(l_j\ra l_i\gamma)=\frac{48\pi^3\alpha}{G_F^2}
               \left((A_M^L)^2+(A_M^R)^2\right)
\]

Our approach to determine the mixing effects is the following:
\begin{itemize}
\item
{}For each particular model we construct the lepton and left and
right slepton mass matrices and determine the corresponding 
diagonalizing matrices.
\item
{}From the input parameters at  $M_U$  ($\alpha_U,\,M_U,\, m_{1/2}$,
e.t.c.) and using the RGEs, we determine the soft masses for gauginos and
sleptons, the Yukawa couplings $\lambda_{t}$ and $\lambda_{b}$ and
the $\mu$ and $A$ parameters. We are using  two-loop-$\beta$
functions and incorporate threshold effects for the scalars and
gauginos. In Table \ref{results} we show the results of the RGEs
running for four characteristic cases (In the runnings we have
assumed the condition $m_{1/2}>m_{3/2}$. We have not explored the
parameter space where the inverse relation holds).
\item
Using the above values and the form of the mass matrices for gauginos
and sleptons we determine the corresponding mass eigenvalues and
eigenstates. The diagonalizing matrices determined in the first step
are used to transform the slepton mass matrices in the desired basis
where the lepton mass matrix is diagonal.
\item
Having all relevant diagonalizing matrices (for charginos, neutralinos
and sleptons), plus the eigenvalues of the gauginos and sleptons we can 
construct the amplitude of the process.
\end{itemize} 

\begin{table}[h]
\centering
\begin{tabular}{||c||c|c|c|c||}
\hline\hline
Case &(a) &(b) &  (c)  &  (d)
\\
\hline
\multicolumn{5}{||c||}{INPUTS}
\\
\hline\hline
$\alpha_U^{-1},\, M_U/10^{16}$&
24.88,\, 1.0&
25.07,\,0.67&
25.00,\,0.67&
25.00,\,0.67
\\
$m_{3/2},\, m_{1/2}$&
138,\, 350&
150,\, 380&
160,\, 390&
160,\,390
\\
$\lambda_t,\, \lambda_b$&
0.85,\, 0.10&
0.90,\, 0.10&
1.05,\, 0.08&
1.05,\, 0.05
\\
$A,\,\mu$&
-207,\,-395 &
-225,\,-395 &
-240,\,-395 &
-240,\,-395
\\
\hline\hline
\multicolumn{5}{||c||}{RESULTS}
\\
\hline\hline
$M_1,\, M_2,\, M_3$&
151,\, 295,\, 820&
166,\, 321,\, 865&
171,\, 329,\, 888&
170,\, 329,\, 890
\\
$\tilde m^{L}_{(1,2)},\, \tilde m^{L}_{(3)}$&
281,\, 279&
303,\, 301&
314,\, 313&
316,\, 314
\\
$\tilde m^{l}_{(1,2)},\, \tilde m^{l}_{(3)}$&
193,\, 187&
209,\, 203&
219,\, 215&
219,\, 218
\\
$\tilde m^{Q}_{1,2},\, \tilde m^{Q}_3$&
750,\, 678&
789,\, 712&
811,\, 730&
813,\, 733
\\
$\tilde m^{u}_{1,2},\, \tilde m^{u}_3$&
718,\, 564&
754,\, 589&
775,\, 598&
778,\, 599
\\
$\tilde m^{d}_{1,2},\, \tilde m^{d}_3$&
718,\, 564&
754,\, 589&
775,\, 598&
778,\, 599
\\
$\tilde m^{d}_{1,2},\, \tilde m^{d}_3$&
714,\, 199&
749,\, 219&
771,\, 244&
773,\, 252
\\
$A_{\tau},\, \mu$&
10.3,\, -324&
8.3,\,  -319&
3.3,\,  -299&
6.9,\,  -301
\\
$\tan\beta,\,m_t$&
14,\,185&
14,\,188&
11,\,195&
7, \,194
\\
\hline
\end{tabular}
\caption
{{Inputs and outputs of the RGEs running for four representative
cases (masses are in GeV).}
\label{results}}
\end{table}

We present here the relevant mass matrices of two models whose
successful fermion mass hierarchy is predicted by $U(1)$ symmetries.
As a first example, we use the scalar mass matrices obtained
in a simple $SU(3)\times SU(2)\times U(1)$ model with an
additional $U(1)_f$ symmetry
\cite{IR}. 
After the implementation of this symmetry, the fermion matrix
for leptons in this model is given
\begin{equation}
{m_\ell}\approx \left (
\begin{array}{ccc}
\tilde{\epsilon}^{2|a +b|} &
        \tilde{\epsilon}^{|a|} &
              \tilde{\epsilon}^{|a +b|}
\\
\tilde{\epsilon}^{|a|} &
        \tilde{\epsilon}^{2|b|} &
             \tilde{\epsilon}^{|b|}
\\
\tilde{\epsilon}^{|a +b|} &
       \tilde{\epsilon}^{|b|} &1
\end{array}
\right){m_{\tau}}
\label{ml0}
\end{equation}
where the parameter $\tilde\epsilon$ is some power of the singlet vev scaled
by the unification mass, while $a,b$ are certain combinations of the
lepton and quark $U(1)_f$-charges. Order one parameters in front of the
various entries (not calculable in this simple model) are assumed to
reproduce exactly the fermion mass relations after renormalization
group running.\footnote{We use the Yukawa coefficients
$c_{11}=4.0; c_{12}=c_{21}=0.9; c_{22}=1.08; c_{33}=1.9.$}
  The scalar mass matrices of this model are given in
\cite{LT}.
{}For the sleptons we obtain
\beq
\tilde m^2_{\ell,e_R} \approx
\left (
\begin{array}{ccc}
{1} & \tilde\eps^{\mid a + 6 b\mid  }
&\tilde\eps^{\mid a + b\mid} \\
\tilde\eps^{\mid a + 6 b \mid } & {1} &
\tilde\eps^{\mid b \mid}\\
\tilde\eps^{\mid a + b\mid } & \tilde\eps^{\mid b\mid} & 1
\end{array}
\right)m_{3/2}^2
\eeq
A successful lepton mass hierarchy in this case is obtained
for the choice $a=3, b=1$ and $\tilde\eps = 0.23$.

As a second example in the calculation of the rare processes,
we apply our results in several textures obtained in context
of the $SU(4)$ model in ref.\cite{AKLL}.
As far as the symmetric
fermion mass textures  are concerned,
the analysis follows the same lines as in the first example,
since the matrices obtained
are similar to those above. The non-symmetric case is however
completely different. There, the violations are much harder
and the bounds on the various scalar mass parameters increase
substantially. In order to be specific, we work out a particular
example based on the charge assignment of Table \ref{charges}.
The field assignment is $F+\bar{F}=(4,2,1)+(\bar 4,1,2)$ for
the three generations, $H+\bar H(4,1,2)+(\bar 4,1,2)$
for the higgses and $h=(1,2,2)$ for the bidoublet including the standard
higgs doublets.
\begin{table}[h]
\centering
\begin{tabular}{cccccccccc}\hline
 $F_1$ & $F_2$ & $F_3$ &  $\bar{F}_1$ & $\bar{F}_2$ & 
$\bar{F}_3$ & $h$ &
 $H$ & $\bar{H}$ \\ \hline
4 & 0 & -1 & 3 & -1 & 1 & 0 &  1 & -1 \\
\hline
\end{tabular}
\caption{{\protect\small $U(1)_{f}$ charges of fields in the model.}
\label{charges}}
\end{table}
Notice that this $U(1)$ symmetry is anomalous. However the
mixed anomalies are zero. This fact allows for a Green-Schwarz anomaly
cancellation mechanism. 
The fermion mass matrices are generated by operators of the form
\bea
\bar{F}_i{F}_j h \delta^m\eps^n
\eea
where $\delta = H\bar{H}$ is an effective singlet generated
by the higgs fields $H,\bar{H}$.

 Taking into account  Clebsch-Gordan coefficients derived from
these operators,  as in ref.
\cite{AKLL}
and using the values $\eps\sim 0.14,\;\delta\sim 0.21$,
we arrive at the following lepton Yukawa matrix at the unification scale 
\begin{eqnarray}
\lambda_e &=& \left(\begin{array}{ccc}
0 & \sqrt{2}\eta_{12} & 0\\
\sqrt{2}\eta_{21} & 3\eta_{22}/\sqrt{5} & 0 \\
0  & -3 \sqrt{2}\eta_{32}/\sqrt{5}    & \eta_{33} \\ \end{array}\right)
\label{nonsymleptoncomponents}
\end{eqnarray}
The following unification scale input parameters give the correct
lepton mass spectrum
\begin{eqnarray}
\eta_{22} = 2.88\times 10^{-2}, & \eta_{12} = 2.81\times 10^{-3}, &
\eta_{21} = 1.30 \times 10^{-3}, \nonumber
\\
 \eta_{33} = 1.18,&  
\eta_{32} = 7.28\times 10^{-2}.  &
\label{inputs}
\end{eqnarray}

The scalar mass matrices are determined from the $U(1)$
charges chosen to give correct predictions for the fermions.
Since left and right fields have different charges, we
obtain two different types of scalar mass matrices 
\begin{eqnarray}
\tilde{m}_l^2 &=& \left(\begin{array}{ccc}
1&\bar\eps^4&\bar\eps^5\\
\eps^4&1&\bar\eps\\
\eps^5&\eps&1\\ \end{array}\right)
\label{scaleft}
\\
\tilde{m}_{e_R}^2&=& \left(\begin{array}{ccc}
1&\bar\eps^4&\bar\eps^2\\
\eps^4&1&\eps^2\\
\eps^2&\bar\eps^2&1\\ \end{array}\right)
\label{scalright}
\end{eqnarray}
The two expansion parameters $\epsilon$ and
$\bar\eps $ are defined as follows
\begin{equation}
 \epsilon \equiv \langle \theta\rangle /{M_U} \sim
\bar\eps =\langle \bar{\theta}\rangle /{M_U} 
\end{equation}

In Table \ref{BR} we show the branching ratios for the four cases
appearing in Table \ref{results} and for the two models discussed
above. We also give the $L$ and $R$ amplitudes for the neutralino
and chargino exchanges for each case (see Eqs \ref{general},
\ref{ampl}).

\begin{table}[p]
\centering
\begin{tabular}{||c||c|c|c|c||}
\hline\hline
CASE &  (a) &  (b) &  (c)    &  (d)
\\
\hline
\multicolumn{5}{||c||}{symmetric textures in $U(1)_f$-models.}
\\
\hline\hline
$A_{M(n)}^L        $  &
$3.58\cdot 10^{-11}$&
$2.74\cdot 10^{-11}$&
$1.22\cdot 10^{-11}$&
$1.83\cdot 10^{-12}$
\\
$A_{M(n)}^R        $ &
$-9.47\cdot 10^{-12}$&
$-6.79\cdot 10^{-12}$&
$-2.04\cdot 10^{-12}$&
$-1.11\cdot 10^{-12}$
\\
$ A_{M(c)}^L       $  &
$4.13\cdot 10^{-14}$&
$3.57\cdot 10^{-14}$&
$1.44\cdot 10^{-14}$&
$8.84\cdot 10^{-15}$
\\
$A_{M(c)}^R        $ &
$8.43\cdot 10^{-12}$&
$7.30\cdot 10^{-12}$&
$2.93\cdot 10^{-12}$&
$1.79\cdot 10^{-12}$
\\
\hline
$BR(\mu\rightarrow e\gamma)$&
$1.03\cdot 10^{-10}        $&
$6.01\cdot 10^{-11}        $&
$1.20\cdot 10^{-11}        $&
$3.07\cdot 10^{-13}$
\\
\hline\hline
$A_{M(n)}^L        $ &
$9.15\cdot 10^{-10}$&
$7.05\cdot 10^{-10}$&
$3.23\cdot 10^{-10}$&
$4.98\cdot 10^{-11}$
\\
$A_{M(n)}^R        $ &
$2.45\cdot 10^{-10}$&
$1.83\cdot 10^{-10}$&
$6.71\cdot 10^{-11}$&
$3.18\cdot 10^{-11}$
\\
$A_{M(c)}^L         $&
$-1.36\cdot 10^{-11}$&
$-1.18\cdot 10^{-11}$&
$-4.74\cdot 10^{-12}$&
$-2.91\cdot 10^{-12}$
\\
$A_{M(c)}^R         $&
$-2.26\cdot 10^{-10}$&
$-1.96\cdot 10^{-10}$&
$-7.87\cdot 10^{-11}$&
$-4.80\cdot 10^{-11}$
\\
\hline
$BR(\tau\rightarrow \mu\gamma)$&
$6.50\cdot 10^{-8}            $&
$3.84\cdot 10^{-8}            $&
$8.08\cdot 10^{-9}            $&
$1.96\cdot 10^{-10}$
\\
\hline\hline
\multicolumn{5}{||c||}{Non-symmetric fermion mass textures}
\\
\hline\hline
$A_{M(n)}^L         $&
$-4.55\cdot 10^{-10}$&
$-3.56\cdot 10^{-10}$&
$-1.70\cdot 10^{-10}$&
$-2.55\cdot 10^{-11}$
\\
$A_{M(n)}^R         $&
$ 6.98\cdot 10^{-12}$&
$ 5.30\cdot 10^{-12}$&
$ 1.72\cdot 10^{-12}$&
$ 1.59\cdot 10^{-12}$
\\
$A_{M(c)}^L         $&
$-2.41\cdot 10^{-14}$&
$-5.98\cdot 10^{-14}$&
$-1.22\cdot 10^{-14}$&
$-1.48\cdot 10^{-12}$
\\
$A_{M(c)}^R         $&
$-4.87\cdot 10^{-12}$&
$-1.21\cdot 10^{-11}$&
$-2.43\cdot 10^{-12}$&
$-2.97\cdot 10^{-12}$
\\
\hline
$BR(\mu\rightarrow e\gamma)$&
$1.66\cdot 10^{-8}        $&
$1.01\cdot 10^{-9}        $&
$2.30\cdot 10^{-9}        $&
$5.21\cdot 10^{-11}$
\\
\hline\hline
$A_{M(n)}^L         $&
$-8.51\cdot 10^{-10}$&
$-6.56\cdot 10^{-10}$&
$-3.03\cdot 10^{-10}$&
$-4.39\cdot 10^{-11}$
\\
$A_{M(n)}^R         $&
$ 1.59\cdot 10^{-10}$&
$ 1.16\cdot 10^{-10}$&
$ 3.38\cdot 10^{-11}$&
$ 2.93\cdot 10^{-11}$
\\
$A_{M(c)}^L         $&
$-1.53\cdot 10^{-11}$&
$-1.32\cdot 10^{-11}$&
$-5.31\cdot 10^{-12}$&
$-3.27\cdot 10^{-12}$
\\
$A_{M(c)}^R         $&
$-2.55\cdot 10^{-10}$&
$-2.21\cdot 10^{-10}$&
$-8.86\cdot 10^{-11}$&
$-5.40\cdot 10^{-11}$
\\
\hline
$BR(\tau\rightarrow \mu\gamma)$&
$6.08\cdot 10^{-8}            $&
$3.67\cdot 10^{-8}            $&
$7.83\cdot 10^{-9}            $&
$2.27\cdot 10^{-10}$
\\
\hline\hline
\end{tabular}
\caption
{Branching ratios for the two processes $\mu\rightarrow e\gamma$
and $\tau\rightarrow\mu\gamma$,
corresponding to the four cases of Table 1
and for the two models with symmetric and non-symmetric
mass textures.
We also show the L and R amplitudes with neutralino and
chargino exchanges.\label{BR}}
\end{table}

{}From Tables \ref{results} and \ref{BR}, we can conclude that lepton
flavour violation in a viable class of supersymmetric unified theories
puts non-trivial constraints on the scalar mass spectrum. In the
case of symmetric textures of mass matrices, the
 $\mu\ra e\gamma$-branching ratio is found to exceed the present 
experimental bound ($4.9\times 10^{-11}$) for values of the 
gravitino mass parameter $m_{3/2}$ less than about $150$Gev, 
in particular when $\tan\beta$ obtains intermediate or higher
values.  
In our approach, the gaugino mass parameter $m_{1/2}$  is also
found to have a lower bound ($\ge 350$Gev) for consistency with
the experimental $\mu\ra e\gamma$-bound. In the case of non- 
symmetric masses matrices, the bounds are even higher: the 
three cases $(a,b,c)$ of Table 3 violate the present experimental
bounds, implying therefore a rather heavy sypersymmetric mass spectrum. 
{F}or small $\tan\beta$, however, we obtain results consistent with
the experimental bounds as in case $(d)$.
Actually, we see a strong dependence of the branching ratio
on $\tan\beta$. As $\tan\beta\ra 1$, the neutralino exchange
processes get smaller and smaller. The chargino ones, while
individually remain of the same order, they exhibit an increasing
cancelation, rendering the branching ratio much lower that the
experimental bound. 
The $\tau\ra\mu\gamma$-rare decay (branching ratio $<4.2\times 10^{-6}$),
does not put further constraints as can been checked from the same
Table. Moreover, the branching ratio for the $\mu\ra 3 e$-decay
in all cases is found about three orders smaller than $BR(\mu\ra e\gamma)$.

Returning to the results of Table 1, we infer that non-observation of
the $\mu\ra e\gamma$-decay implies that all  scalars appear with
masses at least heavier than about $200$Gev. Although our results are only
for two specific examples, their main characteristics
are rather generic and all sparticle mass bounds from flavour decays
are expected always much larger than those obtained from other
types of experiments.   In this selected region of
the parameter space (i.e. $m_{1/2}>m_{3/2}$) the sleptons are
the lightest scalars.  Clearly, the  relatively large flavour violations 
are due the large $\tan\beta$-effects as well as 
to the fact that $U(1)_f$-models imply also mixing effects to the 
scalar sector. The rather high scalar mass bounds could be 
considered as an ominous  perspective for models with large 
 $\tan\beta$ and non-symmetric fermion mass textures, 
in particular for those who envisage a relatively light
supersymmetric mass spectrum accessible to future experiments.
We should note however, that slepton masses of this
order are in the range of the LHC. Indeed, slepton decays can in
principle be detected in CMS with a mass up to $350$Gev\cite{r}
 {}Finally, we wish to comment that in more complicated
structures with cyclic permutation symmetries between generations
and universal anomalous $U(1)$-factors may prevent mixing effects
in the supersymmetric mass matrices\cite{AP}. In such cases, the
above constraints are relaxed.

In conclusion,  we have examined lepton flavour rare processes 
in a class of supersymmetric gauge theories where an additional
$U(1)$ symmetry discriminates the  three families. Such symmetries,
are capable of generating successfully the hierarchical fermion
mass spectrum and the Kobayashi-Maskawa mixing in the hadronic
sector, when non-renormalizable contributions are taken into
account in the superpotential. They, however,
imply mixing effects through non-renormalizable terms in the 
K\"ahler potential, and consequently, in the scalar partners
 of quarks and leptons, leading thus to hard flavour
violations. These violations are even larger when models with non-symmetric 
textures and large $\tan\beta$ values are considered.
  As a result, stringent constraints are found in the
sparticle spectrum.
 Such bounds could be relaxed if, for example,
additional symmetries of the K\"ahler potential are imposed,
so that the scalar partners of fermions remain flavour diagonal,
while at the same time they do not appear in the superpotential. 
We hope to come back to this issue in a future publication.


\end{document}